\documentclass[12pt]{article}
\pdfoutput=1
\usepackage{amstext,amsmath,amssymb,amsfonts}
\usepackage{bbm}
\usepackage[latin1]{inputenc}
\usepackage{hyperref}
\usepackage{amsthm}
\usepackage{color}
\usepackage{geometry}
\usepackage{graphicx}
\usepackage{framed}
 
\newtheorem{definition}{Definition}

\newtheorem{proposition}{Proposition}
 
\newcommand{\Tr}{\mathrm{Tr}}
 
\newcommand{\be}{\begin{equation}}
\newcommand{\ee}{\end{equation}}
  
\newcommand{\bea}{\begin{eqnarray}}
\newcommand{\eea}{\end{eqnarray}}

\begin{document}
\begin{flushright}
 IPHT-T14/142 \\
 CRM2014-3342
\end{flushright}

\vspace{20pt}

\begin{center}

{\Large\bf 
An analysis of the intermediate field theory of $T^4$ tensor model.}
\vspace{20pt}

Viet Anh Nguyen${}^{a,c,e}$, St\'ephane Dartois${}^{a,b}$, Bertrand Eynard${}^{c,d,\ddagger}$

\vspace{10pt}

\begin{abstract}
\noindent In this paper we analyze the multi-matrix model arising from the intermediate field representation of the tensor model with all quartic melonic interactions. We derive the saddle point equation and the Schwinger-Dyson constraints. We then use them to describe the leading and next-to-leading eigenvalues distribution of the matrices.
\end{abstract}
\end{center}


\setcounter{footnote}{0}
\setcounter{lemma}{0}
\setcounter{theorem}{0}

\section{Introduction}

 Tensor models have been introduced as a generalization of matrix models. They were first presented in \cite{oldgft1,oldgft2} in order to give a description of quantum gravity in dimension $D> 2$ as a field theory \emph{of} space-time (and not \emph{on} space-time). To this aim they were really inspired by matrix models. Indeed the field theory thus obtained generated Feynman graphs that may have\footnote{up to some conventional added informations.} an interpretation as a $D$-dimensional space (but this space was not a manifold in general). Each of these graphs therefore came with a quantum amplitude associated to the field theory Feynman rules. But unfortunately they turned out to be very difficult to handle analytically because of the lack of tools allowing to compute them and the lack of theoretical understanding of what is a tensor and how it should be understood in this context. 
 Moreover the geometry of three and more dimensional spaces is considerably more involved than the $2$-dimensional geometry. This is the source of difficulties when trying to give a combinatorial description of these spaces fitting with the field theory combinatorics. 
 \medskip
 
 On the other hand matrix models were well developed. In fact eigenvalues, characteristic polynomials and determinants are objects allowing to effectively compute quantities of interests and thus to gain understanding of the respective models that one introduced. Also the $1/N$ expansion \cite{'tHooft:1973jz} was a crucial tool to these advances and was lacking in the tensor models framework. This expansion enabled to solve combinatorial problems (for instance \cite{Itzykson:1979fi}) in a beautiful manner. The double scaling limit provided a road to the non-perturbative definition of string theory and thus attracted activities from this area. Moreover the relationship to Liouville's theory of quantum gravity in the continuum is still a problem under consideration keeping the community busy. The integrability and geometric structures unravelled in many of these models allowed to build a rich theory of matrix models and their related geometry. It was related to interesting concepts such as Hirota's equations, orthogonal polynomials, KdV hierarchy, intersection number for moduli spaces, $2D$ topological field theory and Frobenius manifold. It was also the source of the so called Eynard-Orantin Topological Recursion \cite{Eynard:2004mh} that is now applied far beyond the scope of original matrix models \cite{EOAlgRM}, allowing to solve many problems of algebraic and enumerative geometry. This recursion is still lacking a complete comprehensive algebraic geometry formulation but promises to be a fruitful source of new concepts for geometry in the near future.
 \medskip
 
 These remarks being made, \emph{colored} tensor models have been introduced in \cite{color}. Early work on these models showed that many difficulties of the early tensor models were solved in this setting. The most important issue solved by \emph{colored} tensor models was the lack of $1/N$ expansion \cite{expansion3}. Contrary to matrix models this expansion was not topological in a naive way. Indeed the parameter governing this expansion, called the \emph{degree}, is not a  topological invariant of the space corresponding to the Feynman graph. The geometric interpretation of this parameter is still unclear. Fortunately it can be computed rather simply from the combinatorial description of the Feynman graph.
  \medskip
 
 Moreover these new tensor models setting enabled a non-ambiguous description of the observables of the models \cite{uncoloring}. This allowed to make complete computation of the $N\rightarrow \infty$ limit, as well as next-to-leading order computations \cite{critical, KOR}. After that came the computation of the double scaling limit of these models \cite{Dartois:2013sra,GS,Bonzom:2014oua} and the coupling to matter \cite{IsingD,spinglass}. Non-perturbative results were also obtained \cite{Gurau}. From these last works one could notice that a class of simple tensor models can be formulated as (rather complicated) multi-matrix models. In this paper we are interested in analyzing these new matrix models using techniques that are classical in the context of standard matrix models. We compute the eigenvalues distribution of the matrix formulation of the simplest tensor model, the \emph{quartic melonic} model. We study this distribution up to the next-to-leading order (NLO) in $1/N$ using saddle point equations and Schwinger-Dyson equations. 
 
 \medskip
 This paper is organized as follows:
\begin{itemize}
\item Section \ref{sec:Matrix} recalls basic results about matrix models and techniques allowing to make computations in their setting.
\item Section \ref{sec:color} introduced the general setting of colored tensor models. Then the specific tensor model under consideration is defined and its matrix formulation derived. We end this section by establishing Proposition \ref{prop:TensMat}, which gives a simple relation between the observables of the tensor models and the observables of the related matrix model.
\item Section \ref{sec:saddle} is devoted to saddle point computations of the eigenvalues distribution at leading order and next-to-leading order. 
\item Section \ref{sec:S-D} describes the Schwinger-Dyson equations leading to a reformulation of the matrix model that is more suited for computation. The NLO eigenvalues distribution is computed from these equations.
\end{itemize}

\section{Matrix Models}

\label{sec:Matrix}
For pedagogical reasons we review here in the context of simple matrix models the tools we shall use in our study. Everything 
in this section can be found in the literature (for instance see \cite{matrix}).

\subsection{Generalities on Matrix Models.}
\medskip
{\bf Matrix $1/N$ Expansion:} \newline
 We recall briefly the $1/N$ expansion of matrix models. Consider the matrix model defined by:
 \be
  Z[t_4,N]=\int dM \exp\left(-N\bigl( \frac{1}{2}\Tr(M^2) +\frac{t_4}{4} \Tr(M^4)  \bigr)\right),
 \ee
$N$ being the size of the matrix. At the formal level this is a generating function for quadrangulations. The free energy $F=\ln Z$ expands as $F=\sum_{g\ge 0} N^{2-2g} F_g(t_4)$ where the $F_g$'s are generating functions of quadrangulations of genus $g$ for the counting variable $t_4$ for quadrangles. In the limit $N \rightarrow \infty$ only the leading order survives {\it i.e.} $F_0$ which counts the \emph{planar} quadrangulations, hence quadrangulations of the sphere $S^2$.
 One can compute the two point function $G_2(t_4)= \langle \Tr(M^2)\rangle$ in this limit and recover Tutte's result for planar rooted quadrangulations:
 \be 
  G_2(t_4,N=\infty)= \sum_n 2 \frac{3^n}{(n+2)(n+1)} \binom{2n}{n} (-t_4)^n.
 \ee
\medskip
\noindent {\bf Saddle Point Method:} \newline
Let us introduce the Hermitian $1$-matrix model by the partition function
\be
Z_{1MM}[\{t_p\},N]=\int dM \exp(-N(\frac{1}{2} \Tr(M^2) + \sum_{p=0}^d t_p \Tr(M^p))).
\ee
It can be rephrased using eigenvalues variables as
\be Z_{1MM}=\int\prod_{i=1}^N d\lambda_i \Delta(\{\lambda_j\})^2\exp(-N(\frac{1}{2}\sum_j \lambda_j^2 + \sum_p^d t_p  \sum_j \lambda^p)),\ee
$\Delta$ being the Vandermonde determinant. 

\medskip
The integrand can be rewritten as $\exp(-N^2 S(\{\lambda_k\}))$ by taking the logarithm of $\Delta(\{\lambda_j\})^2 = \exp(\log(\Delta(\{\lambda_j\})^2))$. The saddle point approximation is given by the value of the integrand on its extrema. Looking for such extrema leads to the equation
\bea
&0=&\frac{1}{N}\lambda_{\nu} + \frac{1}{N}V'(\lambda_{\nu})- \frac{1}{N^2} \sum_{i \neq \nu} \frac{1}{\lambda_{\nu}-\lambda_i}.
\eea
This equation can be solved in the $N \rightarrow \infty$ limit by introducing the resolvent $W(x) = \sum_i \frac{1}{x-\lambda_i}$. In fact one has the following well known relation (re-demonstrated later in this paper):
\be
W(x)^2 = \frac{1}{N^2} \sum_{k,j|k \neq j} \Bigl(\frac{1}{x-\lambda_k}-\frac{1}{x-\lambda_j}\Bigr)\frac{1}{\lambda_k-\lambda_j} - \frac{1}{N} W'(x).
\ee
But the first term of the RHS can be computed from the saddle point equation, giving\footnote{Defining $V(x)=\sum_{p=0}^d t_p x^p$.} 
\be
W(x)^2=\frac{2}{N}\sum_k\frac{\lambda_k + V'(\lambda_k)}{x-\lambda_k} -\frac{1}{N}W'(x),
\ee
and so 
\be
W(x)^2= 2(x+ V'(x))W(x) -\frac{1}{N} W'(x) -2 +\sum_k \frac{V'(\lambda_k) - V'(x)}{x-\lambda_k} .
\ee
Actually the last term is a polynomial, called $P(x)$, because $V'$ is a polynomial and $V'(\lambda_k)-V'(x)$ is a polynomial vanishing at $x=\lambda_k$.
At leading order in $N$ the second term of the RHS is irrelevant and this equation can be solved algebraically. This is what is done later in a more involved case to compute 
the NLO distribution of the matrix formulation of the quartic tensor model considered in this paper.
\newline

\medskip
\noindent{\bf Schwinger-Dyson Constraints and Loop Equations:} \newline
Schwinger-Dyson constraints (or equations) are just relations between correlation functions coming from integration by parts. They are often derived by setting that the integration of a total derivative should be zero. That is the method we use here:
\bea
&0=&\int dM \frac{\partial}{\partial M_{ij}}\Bigl( (M^{k+1})_{ij} \exp(-N (\frac{1}{2} \Tr(M^2) +V(M)))\Bigr)\nonumber \\
&\Leftrightarrow& \nonumber \\
&0=& \frac{1}{N}\langle \sum_{n=0}^{k} \Tr(M^n)\Tr(M^{k-n}) \rangle -\langle \Tr(M^{k+2})\rangle -\langle\Tr(M^{k+1}V'(M))\rangle.
\eea
Multiplying for each values of $k$ by $z^{-k-2}$ and summing over $k$:
\bea
&0=&W(z,z) + W(z)^2-(z+V'(z))W(z) - P(z),
\eea
$P(z)$ being a polynomial. $W(z)$ and $W(z,z)$ are respectively the resolvent (introduced above) and the bi-resolvent. 
\section{Tensor Models} \label{sec:color}

In this section we introduce briefly the general framework of tensor models. 
More details can be found in general references on the subject, for instance the necessary background is contained in \cite{expansion3, uncoloring}.  

\subsection{Tensor Invariants and Generic $1$-Tensor Models}

We construct tensor models in a way similar to the construction of the prototype example of matrix models \textsl{i.e.} the Hermitian one matrix model. The action of this model is constructed as a sum over the $GL(N)$\footnote{acting in an natural way.} invariants of the matrix, in fact one is led to such a choice in order to get a well-defined action over the matrices (and not just arrays of numbers).

Consider a rank $D$ tensor $T$ and its complex conjugate $\bar{T}$ (\textsl{i.e.} a tensor with complex conjugated entries, once a particular choice of basis has been made). In fact $T$ belongs to a space of the form $V_1\otimes \cdots \otimes V_D$ endowed with a Hermitian product and $\bar{T}$ belongs to the canonical dual $\check{V}_1\otimes \cdots \otimes \check{V}_D$.  Consider their components $T_{i_1\cdots i_D}$, $\bar{T}_{i_1 \cdots i_D}$ in a basis. Require for simplicity that $\dim V_j = N$ for all $j$. 
The tensor model should be invariant under the action of $GL(V_1)\times\cdots \times GL(V_D)$. This action can be written explicitly on the components of the tensor $T$: 
\begin{equation}
T_{j_1\cdots j_D}'=R(g_1)_{j_1i_1}\cdots R(g_D)_{j_D i_D} T_{i_1 \cdots i_D}.
\end{equation}
In the dual vector space the action of $GL(V_1)\times\cdots \times GL(V_D)$ is given by:
\begin{equation}
\bar{T}_{j_1\cdots j_D}'=R(g_1)^{-1}_{j_1i_1}\cdots R(g_D)^{-1}_{j_D i_D} \bar{T}_{i_1 \cdots i_D}.
\end{equation}
Thus one can find all the polynomial invariants. They are all obtained by contracting the $j^{th}$ index of a $T$ with the $j^{th}$ index of a $\bar{T}$, in which case all the $R$ 
and $R^{-1}$ matrices cancel out\footnote{Actually this is the only meaningful thing to do in the mathematical setting given here, although it can be easily extended to support contraction of index which are not in the same position.}. So one obtains the tensor invariants as some objects $T\bar{T} T \bar{T} \cdots T\bar{T}$ with a contraction pattern between them that respects the position. Such invariants will be called \emph{trace invariants}.

\begin{figure}
 \begin{center}
  \includegraphics[scale=1.4]{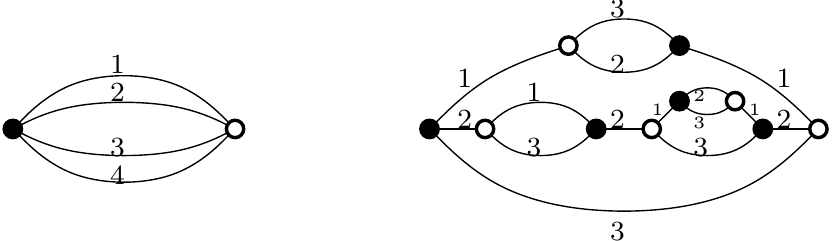}
  \caption{On the left the $4$-colored fundamental melon, the set of colors $\mathcal{C}=\{1,2,3,4\}$. On the right, an example of a $3$-colored melonic graph, here $\mathcal{C}=\{1,2,3\}$.}
  \label{fig:exmelon}
 \end{center}
\end{figure}

They can be graphically represented by $D$-edge-colored bipartite graphs
(hence the name \emph{colored} tensor models). A $D$-edge-colored bipartite graph is a graph
with $v$ black vertices (standing for $T$) and $v$ white vertices (standing for $\bar T$) such
that only vertices of different colors are connected by edges and exactly $D$ edges of $D$ different
colors are attached to each vertex. The color of an edge indicates the position
of the index being contracted. We draw some examples of such graphs for $D=3$ in Fig.\ref{fig:exmelon}.
In particular, the melonic quartic (\textsl{i.e.} with two $T$'s and two $\bar{T}$'s) invariants are represented by three graphs like the second one (with permutations of colors).

For this colored graphs we need to define the \textsl{jackets} and the \textsl{degree}:
\begin{definition}
A colored jacket $\mathcal{J}$ is an edge-colored ribbon graph associated to a $D$-colored graph $\mathcal{G}$ with as 1-skeleton the graph $\mathcal{G}$ and 
with faces made of graph cycles of colors $(\tau^q (0), \tau^{q+1} (0))$ for some cyclic permutation $\tau$ of $D$ elements (the colors), modulo the orientation of the cycle (\text{i.e.} $\tau^{-1}$ leads to the same jacket).  
\end{definition}

As such there are $\frac{(D-1)!}{2}$ jackets for a $D$-colored graph. 
Each jacket $\mathcal{J}$ leads to a cellular decomposition of a surface and thus comes with a genus $g_{\mathcal{J}}$.

\begin{definition}
 The degree of a colored graph $\mathcal{G}$ is:
 \begin{equation}
  \omega(\mathcal{G})= \sum_{\mathcal{J}(\mathcal{G})} g_{\mathcal{J}},
 \end{equation}
hence the sum of the genera of the jackets associated to the graph.
\end{definition}
For $D=3$ for example the degree reduces to the genus of the only jacket associated to the graph. This allows us to define the generic tensor model:
\begin{definition}
The $(D+1)$-dimensional generic tensor model is defined by the partition function:
\bea
Z[N,\{t_{\mathcal{B}}\}]= \int dT d\bar{T} \exp\Bigl(-N^{D-1} \sum_{\mathcal{B}}N^{-\frac{2}{(D-2)!}\omega(\mathcal{B})} t_{\mathcal{B}}\mathcal{B}(T, \bar{T})\Bigr),
\eea
where $\mathcal{B}$ runs over the regular $D$-colored graphs indexing the invariants. \\
The $t_{\mathcal{B}}$ are the coupling constants, the one corresponding to the only invariant of order $2$ often being fixed to $1/2$.
$\mathcal{B}(\cdot,\cdot)$ is the invariant of $T$ and $\bar{T}$ indexed by the graph $\mathcal{B}$. $\omega$ is the degree of the graph $\mathcal{B}$.
\end{definition}

\begin{definition}
 A $D$-colored graph $\mathcal{G}$ is said to be melonic if and only if $\omega(\mathcal{G})=0$.
\end{definition}

The reason of this name becomes transparent when the structure of a melonic graph is described. In  \cite{critical, expansion3} it is shown that all melonic graphs are obtained by recursive insertions of $(D-1)$-dipoles on lines of the fundamental melon (see Fig. \ref{fig:exmelon}).

\subsection{$T^4$ melonic tensor models and intermediate field representation.}

In this section we introduce the model we shall study and we give its representation in terms of matrix integrals.

\begin{figure}
 \begin{center}
  \includegraphics[scale=0.8]{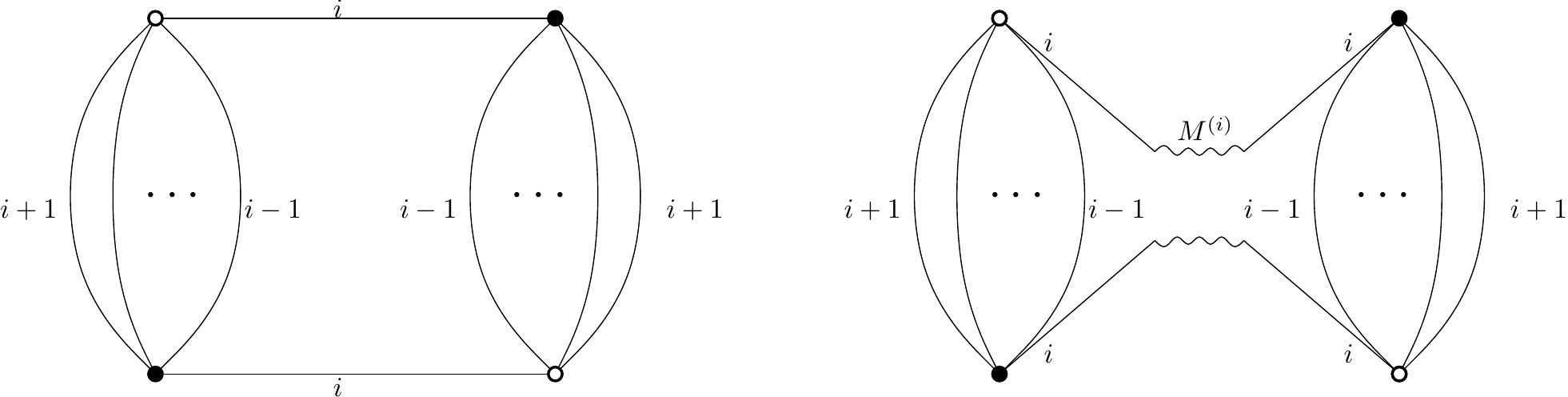}
  \caption{On the left one of the interaction term. On the right its splitting with the intermediate matrix field of color $i$: $M^{(i)}$.}
  \label{fig:intsplit}
 \end{center}
\end{figure}

We study the quartic or $T^4$ melonic tensor model in $D$ dimension, which is the simplest. Its name indicates that we choose as interaction terms the simplest ones {\it i.e.} those that are represented by melonic $D$-colored graphs of the form of Fig. \ref{fig:intsplit}.

In order to write the model let us introduce some notations. We call $\mathcal{C}$ the set of colors, so to say the set on index of the components of the tensor.
We then write $\bar{T} \cdot T$ the contraction of all the indices of $\bar{T}$ with all the indices of $T$. Then we introduce the partial scalar product between $\bar T$ and $T$. Let $\mathcal{Y}\subset \mathcal{C}$ be a subset of $\mathcal{C}$. We denote $\bar{T} \cdot_{\mathcal{Y}} T$ the contraction of indices of $\bar{T}$ which belong to $\mathcal{Y}$ with the indices of $T$ which also belong to $\mathcal{Y}$. Moreover we denote $\hat{\mathcal{Y}}$ the complementary of $\mathcal{Y}$ into $\mathcal{C}$, if $\mathcal{Y}$ reduces to one element, we denote it by the element.

The partition function of our model is given by:
\bea
Z = \int_{(\mathbb{C}^{N})^{\otimes D}} dT d\bar{T} \exp\Biggl(-N^{D-1}\biggl(\frac{1}{2}(\bar{T}\cdot T) +\frac{\lambda}{4}\sum_c 
(\bar{T}\cdot_{\hat{c}} T)\cdot_c (\bar{T}\cdot_{\hat{c}} T)\biggr)\Biggr).
\eea
We introduce an intermediate matrix field $M^{(c)}$ used to split the interaction terms $(\bar{T}\cdot_{\hat{c}} T)\cdot_c (\bar{T}\cdot_{\hat{c}} T)$. This is pictured on the right of Fig. \ref{fig:intsplit}. Doing this allows one to construct a matrix model that is equivalent to the tensor model under consideration.
We write the interaction term as:
\bea
&\exp\Bigl( -N^{D-1}\frac{\lambda}{4}(\bar{T}\cdot_{\hat{c}} T)\cdot_c (\bar{T}\cdot_{\hat{c}} T)\Bigr)  = \nonumber \\ 
&\int dM^{(c)}\exp(-\frac{N^{D-1}}{2}
\Tr((M^{(c)})^2)-i \sqrt{\lambda/2} N^{D-1}\Tr((\bar{T}\cdot_{\hat{c}} T) M^{(c)}).
\eea
This choice of scaling for the matrix allows to suppress the factor of $N$ in the logarithmic potential that one is going to obtain after integrating out the tensor degrees of freedom. 
Rewriting the tensor model using this representation of the interaction term we get:
\bea
\label{eq:mixed}
Z&& = \int_{(\mathbb{C}^{N})^{\otimes D}} dT d\bar{T} \int_{H_N^D} \prod_c dM^{(c)}\nonumber \\
&&\exp\Biggl(-\frac{N^{D-1}}{2} \bar{T}\biggl(\mathbbm{1}^{\otimes D} +i \sqrt{\lambda/2}\sum_{k=1}^D \mathcal{M}_k\biggl) T\Biggr) \exp\Biggl(-\frac{1}{2}\sum_m \Tr(\mathcal{M}_m^2))\Biggr), \nonumber
\eea
where we introduced the notation $\mathcal{M}_m = \mathbbm{1}^{\otimes(m-1)}\otimes M^{(m)} \otimes\mathbbm{1}^{\otimes (D-m)}$ for any $m\in [\![ 1, D ]\!]$. Integrating out the $T$'s we obtain:
\bea
Z&& = \int_{(H_N)^D} \prod_c dM^{(c)} \det {}^{-1}\Bigl(\mathbbm{1}^{\otimes D} +i \sqrt{\lambda/2}\sum_{k=1}^D \mathcal{M}_k\Bigr)\exp\Biggl(-\frac{1}{2}\sum_m \Tr(\mathcal{M}_m^2))\Biggr) \nonumber,
\eea
this is the intermediate field representation of the $T^4$ melonic tensor model. 

There are simple relations between the observables of this matrix model and some of the observables of the related tensor model.
\begin{framed}
\begin{proposition} \label{prop:TensMat}
We have: 
\be
\langle\Tr(\Theta_c^p)\rangle=\Bigl(\frac{2i\sqrt{2}}{\sqrt{\lambda}}\Bigr)^p \langle\Tr H_p(M^{(c)})\rangle,
\ee
and
\be
\langle\Tr(M^{(c)}{}^p)\rangle= \langle\Tr H_p(\frac{\sqrt{\lambda}}{2i\sqrt{2}} \Theta_c)\rangle,
\ee
where $\Theta_c = (\bar{T}\cdot_{\hat{c}} T)$ is a matrix, and $H_p$ is the Hermite polynomial of order $p$.
\end{proposition}
\end{framed}
{\bf Proof:}
Consider the mixed matrix-tensor representation of \ref{eq:mixed}. One can write $\langle\Tr(\Theta_c^p)\rangle$ as:
\bea
&&\Bigl(\frac{N^{D-1} \sqrt{\lambda/2}}{2i} \Bigr)^p \langle\Tr(\Theta_c^p)\rangle = \frac{1}{Z}\int_{(\mathbb{C}^{N})^{\otimes D}} dT d\bar{T} \int_{(H_N)^D} \prod_c dM^{(c)}\nonumber \\
&&\Biggl(\frac{\partial^p}{\partial M^{(c)}_{a_1a_2} \partial M^{(c)}_{a_2 a_3} \cdots \partial M^{(c)}_{a_p a_1}}\exp\Biggl(-\frac{N^{D-1}}{2} \bar{T}\biggl(\mathbbm{1}^{\otimes D} +i \sqrt{\lambda/2}\sum_{k=1}^D \mathcal{M}_k\biggl) T\Biggr) \Biggr) \nonumber \\ &&\exp\Biggl(-\frac{1}{2}\sum_m \Tr(\mathcal{M}_m^2))\Biggr),
\eea
with the convention that repeated indices are summed
\footnote{The factor $1/Z$ is generally omitted in the next computation since it is not relevant.}. Up to integration by parts:
\bea
\Bigl(\frac{-iN^{D-1} \sqrt{\lambda/2}}{2} \Bigr)^p \langle\Tr(\Theta_c^p)\rangle = (-1)^p\int_{(\mathbb{C}^{N})^{\otimes D}} dT d\bar{T} \int_{(H_N)^D} \prod_c dM^{(c)}\nonumber \\
\exp\Biggl(-\frac{N^{D-1}}{2} \bar{T}\biggl(\mathbbm{1}^{\otimes D} +i \sqrt{\lambda/2}\sum_{k=1}^D \mathcal{M}_k\biggl) T\Biggr) 
\nonumber \\
\Biggl(\frac{\partial^p}{\partial M^{(c)}_{a_1a_2} \partial M^{(c)}_{a_2 a_3} \cdots \partial M^{(c)}_{a_p a_1}}\exp\Biggl(-\frac{1}{2}\sum_m \Tr(\mathcal{M}_m^2) \Biggr)\Biggr).
\eea
Recall the definition of Hermite polynomials $H_q(x) = (-1)^q\exp(\frac{x^2}{2}) \frac{d^p}{dx^p} \exp(-\frac{x^2}{2})$. This leads to:
\bea
\Bigl(\frac{-iN^{D-1} \sqrt{\lambda/2}}{2} \Bigr)^p \langle\Tr(\Theta_c^p)\rangle =N^{p(D-1)} \langle H_p(M^{(c)})\rangle.
\eea 
For the second equation it suffices to use the Weierstrass transform. It is defined as the linear operator sending a monomial of degree $n$  to the corresponding Hermite polynomial $H_n$. Explicitly we have: 
\be
H_n(x) = e^{-\frac{1}{4}\frac{d^2}{dx^2}} x^n, \forall x\in \mathbb{R}.
\ee
Inverting the operator and using the property of Hermite polynomials $\frac{d}{dx} H_n(x) = nH_{n-1}(x)$ we get:
\be
x^n=\sum_{k=0}^{[n/2]} \frac{1}{4^k} \frac{n!}{(n-2k)! k!} H_{n-2k}(x).
\ee
This can be further used to obtain:
\be
\langle\Tr(M^{(c)}{}^n)\rangle =\sum_{k=0}^{[n/2]} \frac{1}{4^k} \frac{n!}{(n-2k)! k!}\Bigl(\frac{\sqrt{\lambda}}{2i\sqrt{2}}\Bigr)^{n-2k} \langle \Tr(\Theta_c^{n-2k})\rangle,
\ee
hence $\langle \Tr(M^{(c)}{}^n)\rangle= \langle\Tr(H_n(\frac{\sqrt{\lambda}}{2i\sqrt{2}} \Theta_c))\rangle$.
\qed

\section{Saddle Point Equation of the Matrix Model.} \label{sec:saddle}
\subsection{Leading Order (LO) $1/N$ Computation.}
First we write the matrix model in eigenvalues variables:
\bea
Z = \int \prod_{c=1}^D &&\prod_{j=1}^N d\lambda_j^{(c)} \exp\Bigl(-\frac{N^{D-1}}{2}\sum_{c,j} \lambda_j^{(c)}{}^2\Bigr) \nonumber \\ \prod_{\{j_c=1\}_{c=1\cdots D}}^N&& \frac{1}{1+i\sqrt{\lambda/2}\sum_{c=1}^D \lambda^{(c)}_{j_c}}\prod_{c=1}^D \Delta(\{\lambda^{(c)}_j\}_{j=1\cdots N})^2,
\eea
$\Delta$ being the Vandermonde determinant. This can be rewritten as:
\bea
Z=\int \prod_{c=1}^D \prod_{j=1}^N d\lambda_j^{(c)}  \exp(-N^D S(\{\lambda_j^{(c)}\}_{j=1\cdots N}^{c=1\cdots D} )),
\eea   
$S$ being:
\bea
S(\{\lambda_j^{(c)}\}_{j=1\cdots N}^{c=1\cdots D} ) = -\frac{1}{2N}\sum_{c,j} \lambda_j^{(c)}{}^2 &&+ \frac{1}{N^D}\log\Biggl[\prod_{c=1}^D \Delta(\{\lambda^{(c)}_j\}_{j=1\cdots N})^2\Biggr]\nonumber \\  + \frac{1}{N^D} \log &&\Biggl[\prod_{\{j_c=1\}_{c=1\cdots D}}^N \frac{1}{1+i\sqrt{\lambda/2}\sum_{c=1}^D \lambda^{(c)}_{j_c}}\Biggr]. 
\eea
The saddle point equations are given by $\frac{\partial S}{\partial \lambda_k^{(c)}} =0$ for all $(k,c)\in [\![1,N]\!]\times [\![1,D]\!]$.  Thus we obtain the following equations:
\bea
0=&&\frac{\partial S}{\partial \lambda_k^{(c)}} \\  =&& -\frac{\lambda_k^{(c)}}{N} +\frac{1}{N^D} \sum_{l\neq k} \frac{1}{\lambda_k^{(c)}-\lambda_l^{(c)}} -  \frac{i\sqrt{\lambda/2}}{N^D} \sum_{\{j_b\}_{b\neq c}} \frac{1}{1+i\sqrt{\lambda/2}(\lambda_k^{(c)}+\sum_{b\neq c}\lambda_{j_b}^{(b)})}  \nonumber 
\eea
As usual we retrieve the Coulomb potential coming from the Vandermonde determinant. Also the tensor product interaction between the different matrices leads to an interaction term that tends to push all the eigenvalues towards $i\sqrt{\frac{2}{\lambda}}$. Finally the usual Gaussian term tends to attract all the eigenvalues to zero. But this has to be analyzed with care. In fact the scaling in $N$ coming from the tensor model scaling is very different from the one of usual matrix models. 
Since we do not know how to solve these equations exactly we make some hypotheses. First we see that the equations are symmetric under the permutations of the color index $c$. This indicates that the saddle point might obey $\lambda_k^{(c)}= \lambda_k^{(d)}$ for any $c,d=1\cdots D$. So we postulate this property. With this in mind the equations rewrite:
\bea
\label{eq:saddle2}
0= \frac{\lambda_k^{(c)}}{N} -\frac{2}{N^D} \sum_{l\neq k} \frac{1}{\lambda_k^{(c)}-\lambda_l^{(c)}} +  \frac{i\sqrt{\lambda/2}}{N^D} \sum_{\{j_r\}_{r=1\cdots D-1}} \frac{1}{1+i\sqrt{\lambda/2}(\lambda_k^{(c)}+\sum_{r=1}^{D-1}\lambda_{j_r}^{(c)})}.
\eea
Now taking care of the $N$ factors, we see that if we make the hypothesis that $\lambda_k^{(c)}= O(1)$, the first and third terms are leading whereas the second term is a sub-leading $O(\frac{1}{N^{D-2}})$ term, by simple counting arguments.
This motivates an Ansatz (that is checked later) for the expansion of $\lambda_k^{(c)}$ in $1/N$, $\lambda _k^{(c)} = \lambda_{k,0}^{(c)} + \frac{\lambda_{k,1}^{(c)}}{\sqrt{N^{(D-2)}}} +  \frac{\lambda_{k,2}^{(c)}}{N^{(D-2)}} +\cdots$.
We compute $\lambda_{k,0}^{(c)}=\alpha$. In fact the formulation of the matrix model in terms of eigenvalues is totally symmetric with respect to the exchange of these eigenvalues. Thus it should not depend on either $k$ or $c$. We can neglect the second term and we obtain:
\be
\alpha_{\pm}= \frac{-1\pm\sqrt{1+2D\lambda}}{2iD\sqrt{\lambda/2}}.
\ee
We choose the $'+'$ root in order to avoid singularities in the contour of integration. Hereafter it is simply denoted $\alpha$. We obtain:
\begin{framed}
\begin{proposition}
The partition function $Z$ at saddle point is given by $\exp(-N^DS_{\text{saddle}})$:
\be 
Z=(1+2D\lambda)^{N^D/2} \exp\bigl(-\frac{N^D}{4D\lambda}(1+2D\lambda -2 \sqrt{1+2D\lambda})\bigr),
\ee
moreover, the free energy $F=-\log Z$, is given as:
\be
N^D\Bigl(1+2D\lambda -2\sqrt{1+2D\lambda} +1\Bigr) -\frac{1}{2} \log(1+2D\lambda)\Bigr)
\ee
\be
N^D\Bigl(\frac{1}{4D\Lambda}(1+2D\lambda -2\sqrt{1+2D\lambda} +1) -\frac{1}{2} \log(1+2D\lambda)\Bigr)
\ee
\end{proposition}
\end{framed}
{\bf Proof:} Straightforward.

We also get the 2-point function of the tensor model. 

\begin{framed}
\begin{proposition}
The 2-point function $G_2(\lambda) = \frac{1}{N}<\bar{T}\cdot T>$ is given in the $N \rightarrow \infty$ limit by:
\be 
\lim_{N \rightarrow \infty} G_2(\lambda) = \frac{1}{N}<\bar{T}\cdot T>= \frac{1}{N}<\Tr\Theta_c>= \frac{2i\sqrt{2}}{\sqrt{\lambda}}\alpha =\frac{2}{D\lambda} (-1+\sqrt{1+2D\lambda}).
\ee
\end{proposition}
\end{framed}

{\bf Proof:}
Recall the relation of Proposition \ref{prop:TensMat} $<\Tr (\Theta_c^p)>=\Bigl(\frac{2i\sqrt{2}}{\sqrt{\lambda}}\Bigr)^p <\Tr (H_p(M^{(c)}))>$. In the $N \rightarrow \infty$ limit we can compute $<\Tr(M^{(c)})>$ at the saddle point approximation as $\sum_j \lambda_j^c= N a$, thus within this approximation we get $<\Tr (\Theta_c^1)>= \frac{2i\sqrt{2}}{\sqrt{\lambda}} N \alpha$. But $<\bar{T}\cdot T>=<\Tr(\Theta_c)>$ for an arbitrary $c\in[\![1,D]\!]$. Moreover $H_1(x)=x$, from which we deduce the result. Note that it is easy to compute all the $\Tr (\Theta_c^p)$'s in this approximation. \qed

\subsection{Next-to-Leading Order (NLO) Computation.}

In this section we want to compute $\lambda_{j,1}^c$. In particular we see that it has interesting statistical distribution properties. Taylor expanding the saddle point equations \ref{eq:saddle2} in $1/N$, we obtain the following equation for $\lambda_{j,1}^{(c)}$:
\be \label{eq:NLOsaddle}
0=(1-\alpha^2)\lambda_{k,1}^{(c)} - \frac{2}{N}\sum_{l\neq k} \frac{1}{\lambda_{k,1}^{(c)}-\lambda_{l,1}^{(c)}} .
\ee
In fact we have $\lambda_k^{(c)}=\alpha + \frac{\lambda_{k,1}^{(c)}}{\sqrt{N^{D-2}}} + O(\frac{1}{N^{D-2}})$. Inserting this in eq. \ref{eq:saddle2} we get:
\bea
0=\frac{\alpha}{N} &+& \frac{\lambda_{k,1}^{(c)}}{\sqrt{N^D}}- \frac{2}{\sqrt{N^{D+2}}}\sum_{k\neq l} \frac{1}{\lambda_{k,1}^{(c)}-\lambda_{l,1}^{(c)}}  \\&+&\frac{i\sqrt{\lambda/2}}{N^D}\sum_{\{j_r\}}\frac{i\sqrt{\lambda/2}}{1+i\sqrt{\lambda/2} D\alpha+\frac{i\sqrt{\lambda/2}}{\sqrt{N^{D-2}}}(\lambda_{k,1}^{(c)} + \sum_r \lambda_{j_r,1}^{(c)}+O(1/N^{D-2}))}.
\nonumber
\eea
Keeping the relevant terms in $1/N$ and factoring some of them out leads to
\be \label{eq:NLOintermediate}
0=\lambda_{k,1}^{(c)}(1-\alpha^2)-\frac{2}{N}\sum_{l\neq k}\frac{1}{\lambda_{k,1}^{(c)}-\lambda_{l,1}^{(c)}}-\frac{\alpha^2}{N^{D-1}}\sum_{\{j_r\}}(D-1)\lambda_{j_1,1}^{(c)}.
\ee
Summing over $k$ simplifies this equation. By antisymmetry of the Vandermonde factor:
\be
0=(1-D)\alpha^2\sum_k\lambda_{k,1}^{(c)} \Rightarrow \sum_k \lambda_{k,1}^{(c)} = 0. 
\ee
Plugging this into eq. \ref{eq:NLOintermediate} we further obtain:
\be\label{eq:Wigner}
(1-\alpha^2)\lambda_{k,1}^{(c)}-\frac{2}{N}\sum_{l\neq k} \frac{1}{\lambda_{k,1}^{(c)}-\lambda_{l,1}^{(c)}}=0,
\ee
which is the well known equation of the Wigner's semi-circle law. In order to solve it we introduce the (colored) resolvent for the NLO eigenvalues $W_c(x)=\frac{1}{N}\sum_k \frac{1}{x-\lambda^{(c)}_{k,1}}$. Moreover we note that:
\bea
\sum_{k,j|k\neq j}\Bigl(\frac{1}{x-\lambda^{(c)}_{k,1}}-\frac{1}{x-\lambda_{j,1}^{(c)}}\Bigr)\frac{1}{\lambda_{k,1}^{(c)}-\lambda_{j,1}^{(c)}} &=& \sum_{k,j|k\neq j} \frac{1}{(x-\lambda_{k,1}^{(c)})(x-\lambda_{j,1}^{(c)})} \nonumber \\ &=& N^2W_c(x)^2+NW_c'(x)
\eea
and
\bea
\sum_{k,j|k\neq j}\Bigl(\frac{1}{x-\lambda^{(c)}_{k,1}}-\frac{1}{x-\lambda_{j,1}^{(c)}}\Bigr)\frac{1}{\lambda_{k,1}^{(c)}-\lambda_{j,1}^{(c)}}  = 2\sum_{k,j|k\neq j}\frac{1}{x-\lambda^{(c)}_{k,1}} \frac{1}{\lambda_{k,1}^{(c)}-\lambda_{j,1}^{(c)}} .
\eea
The sum over $j$ can be computed from the NLO saddle point equation eq. \ref{eq:Wigner},
\bea
\sum_{k,j|k\neq j}\Bigl(\frac{1}{x-\lambda^{(c)}_{k,1}}-\frac{1}{x-\lambda_{j,1}^{(c)}}\Bigr)\frac{1}{\lambda_{k,1}^{(c)}-\lambda_{j,1}^{(c)}} = \frac{1}{2}\sum_k\frac{(1-\alpha^2)\lambda_{k,1}}{x-\lambda_{k,1}}.
\eea
Thus
\bea
W_c(x)^2= \frac{1}{N}\sum_k\frac{(1-\alpha^2)\lambda_{k,1}}{x-\lambda_{k,1}}-\frac{1}{N}W'(x).
\eea
Since we only consider the $N \rightarrow \infty$ limit, the second term is subleading
\bea
W_c(x)^2 &=& \frac{(1-\alpha^2)}{N}\sum_k \frac{x-(x-\lambda_{k,1})}{x-\lambda_{k,1}} \nonumber \\
&=&(1-\alpha^2)(xW_c(x)-1).
\eea
Hence
\be
W_{c,\pm}(x)= (1-\alpha^2)\biggl(x\pm \sqrt{x^2-\frac{1}{(1-\alpha^2)}}\biggr).
\ee
One notices that the NLO term for the $2$-point function vanishes in this context. Indeed the resolvent is the generating function of the traces of the matrix and the term in front of $1/x^2$ is vanishing in the expansion of $W_{c,-}$. From these two last sections we get the following result:
\begin{framed}
\begin{proposition}
The total resolvent $\mathcal{W}(x)$ of a matrix of any color $c \in [\![1,D]\!]$ expands, up to next-to-leading order, as:
\begin{equation}
\mathcal{W}(x) = \frac{1}{x-\alpha} + \frac{1}{\sqrt{N^{D-2}}} (1-\alpha^2)\biggl(x\pm \sqrt{x^2-\frac{1}{(1-\alpha^2)}}\biggr).
\end{equation}
\end{proposition}
\end{framed}

\section{Schwinger-Dyson Equations.} \label{sec:S-D}
In this part we construct the Loop equations for the model and then use them to derive again the results obtained above. As suggested by the above study, we will consider the loop equations in terms of new variables $\tilde{M}^{(c)}$ defined by $M^{(c)}= \alpha \mathbbm{1} + \frac{\tilde{M}^{(c)}}{\sqrt{N^{D-2}}}$. In fact the previous study showed that in the $N \rightarrow \infty$ all the eigenvalues collapse to a point $\alpha$ and the NLO term follows a distribution which is more regular for a matrix model. Coming back to the expression of $Z$ we have:
\bea
Z = \int_{(H_N)^D} \prod_c dM^{(c)}\det {}^{-1}\Bigl(\mathbbm{1}^{\otimes D} +i \sqrt{\lambda/2}\sum_{k=1}^D \mathcal{M}_k\Bigr)\exp\Biggl(-\frac{1}{2}\sum_m \Tr(\mathcal{M}_m^2))\Biggr) \nonumber \\
=\int_{(H_N)^D} \prod_c dM^{(c)}\exp\Biggl(-\frac{1}{2}\sum_m \Tr(\mathcal{M}_m^2))-\Tr \log \Bigl(\mathbbm{1}^{\otimes D} +i \sqrt{\lambda/2}\sum_{k=1}^D \mathcal{M}_k\Bigr)\Biggr).
\eea
After the change of variables we obtain:
\bea \label{eq:shiftmatrix}
&Z= \frac{\exp\bigl(-\frac{N^D}{2}\alpha^2\bigr)}{N^{D-2}}\int_{(H_N)^D} \prod_c d\tilde{M}^{(c)}\exp\Bigl(-\frac{N}{2}\sum_c \Tr\tilde{M}_c^2 - \alpha N^{\frac{D}{2}}\sum_c \Tr\tilde{M}_c\nonumber \\ 
&-\Tr \log\bigl((1+i\sqrt{\lambda/2}\alpha)\mathbbm{1}^{\otimes D}+i\sqrt{\frac{\lambda}{2 N^{D-2}}}\sum_c\tilde{\mathcal{M}}_c  \bigr)\Bigr),
\eea
with the obvious extension of the previous notation: $\tilde{\mathcal{M}}_c = \mathbbm{1}^{\otimes(c-1)}\otimes\tilde{M}_c\otimes \mathbbm{1}^{\otimes(D-c)}$.
We are now ready to compute the Schwinger-Dyson equations of this model in term of the $\tilde{M}$'s matrices,
\bea
&0=&\frac{\exp\bigl(-\frac{N^D}{2}\alpha^2\bigr)}{N^{2(D-1)}(1+i\sqrt{\lambda/2}\alpha)}\sum_{ij}\int\prod_c d\tilde{M}^{(c)}\frac{\partial}{\partial \tilde{M}_{ij}^{(c)}}\Bigl((\tilde{M}^{(c)})_{ij}^k\exp(-\frac{N}{2}\sum_c \Tr\tilde{M}_c^2 \nonumber \\ &&- \alpha N^{\frac{D}{2}}\sum_c \Tr\tilde{M}_c
-\Tr \log\bigl(\mathbbm{1}^{\otimes D}-\frac{\alpha}{N^{(D-2)/2}}\sum_c\tilde{\mathcal{M}}_c  \bigr))\Bigr),
\eea
from which we obtain:
\bea
&0=&\langle\sum_{n=0}^{k-1} \Tr(\tilde{M}^{(c)}{}^n)\Tr(\tilde{M}^{(c)}{}^{k-1-n})\rangle -N\langle\Tr(\tilde{M}^{(c)}{}^{k+1})\rangle\nonumber \\ 
&-&\langle N^{\frac{D}{2}}\alpha\Tr(\tilde{M}^{(c)}{}^k)\rangle +\langle\sum_{p\ge 0}\Bigl(\frac{\alpha}{N^{(D-2)/2}}\Bigr)^{p+1} \nonumber \\ &&\sum_{\{q_i\}_{i=1\cdots D}|\sum_i q_i=p} \binom{p}{q_1,\cdots , q_D}\Bigl(\prod_{i\neq c}\Tr( \tilde{M}^{(i)}{}^{q_i})\Bigr)\Tr(\tilde{M}^{(c)}{}^{q_c+k}) \rangle,
\eea
the third term canceling with the $p=0$ term of the last sum:
\bea
&0=&\langle\sum_{n=0}^{k-1} \Tr(\tilde{M}^{(c)}{}^n)\Tr(\tilde{M}^{(c)}{}^{k-1-n})\rangle -N\langle\Tr(\tilde{M}^{(c)}{}^{k+1})\rangle\nonumber \\ 
&+&\langle\sum_{p\ge 1}\Bigl(\frac{\alpha}{N^{(D-2)/2}}\Bigr)^{p+1} \nonumber \\ &&\sum_{\{q_i\}_{i=1\cdots D}|\sum_i q_i=p} \binom{p}{q_1,\cdots , q_D}\Bigl(\prod_{i\neq c}\Tr( \tilde{M}^{(i)}{}^{q_i})\Bigr)\Tr(\tilde{M}^{(c)}{}^{q_c+k}) \rangle.
\eea
In the last sum of this equation the only leading term at $N\rightarrow \infty$ limit is the $p=1$ term. In this regime the relevant equation writes:
\bea
&0=&\langle\sum_{n=0}^{k-1} \Tr(\tilde{M}^{(c)}{}^n)\Tr(\tilde{M}^{(c)}{}^{k-1-n})\rangle -N\langle\Tr(\tilde{M}^{(c)}{}^{k+1})\rangle\nonumber \\
&+&\langle\alpha^2 N \Tr(\tilde{M}^{(c)}{}^{k+1})\rangle+\langle\alpha^2\Tr(\tilde{M}^{(c)}{}^k)\sum_{j\neq c}\Tr(\tilde{M}^{(j)})\rangle.
\eea
At the $N\rightarrow \infty$ limit the mean values factorize. In fact looking at the Feynman rules for the model of eq.\ref{eq:shiftmatrix} we obtain the following prescription:
\begin{itemize}
\item edges $\rightarrow$ $\frac{1}{N}$
\item faces $\rightarrow$ $N$.
\end{itemize}
The contribution of the vertices of the graph is more involved. Expanding the potential we notice that the linear term of the expansion vanishes with the term $\alpha N^{\frac{D}{2}}\sum_c \Tr\tilde{M}_c$. The remaining term of the expansion can be represented as vertices of Feynman graphs that are themselves made of $k$ fatvertices of different colors $c\in \mathcal{S} \subset [\![1,D]\!]$, $|\mathcal{S}|=k$ for $1\le k \le D$.
Each fatvertex of color $c$ is of valence $p_c\ge 2$. Each of this vertex comes with a factor $N^{\frac{2-D}{2}\sum p_c+(D-k)}$ (where $D\ge3$).
Since we are interested in the $N \rightarrow \infty$ limit we focus on graphs that are made out of 'leading vertices'. These are the ones for which $k=1$ and $p:=p_c=2$ for a given $c$. The factor coming with these vertices is $N$, it is the usual scaling for matrix models. One can extend the argument for $p\ge 2$ and find the scaling for such graphs $G$ with $E$ edges, $F$ faces and $V$ vertices $N^{F-E +\sum_{v\in G}[(2-D)+(p_v-2)\frac{2-D}{2} +(D-1)]}= N^{\chi(G)-(D-2)(E-V)}$, with $\chi(G)$ the Euler characteristic of $G$. The leading graphs are thus the ones for which $(E-V)$ vanishes and $\chi$ is maximum.
Finally this scaling favors at leading order disconnected contributions maximizing $\chi(G)$. Thus the observables factorize:
\be
\langle\Tr(\tilde{M}^{(l)}{}^s)\Tr(\tilde{M}^{(m)}{}^t)\rangle=\langle\Tr(\tilde{M}^{(l)}{}^s)\rangle  \langle\Tr(\tilde{M}^{(m)}{}^t)\rangle+ O(N^{-(D-2)}).
\ee 
Because of the symmetry we assume $\langle \Tr(\tilde{M}^{(c)})\rangle=0$. This leads to:
\bea \label{eq:LeadSD}
0=\langle\sum_{n=0}^{k-1} \Tr(\tilde{M}^{(c)}{}^n)\Tr(\tilde{M}^{(c)}{}^{k-1-n})\rangle -N(1-\alpha^2)\langle\Tr(\tilde{M}^{(c)}{}^{k+1})\rangle.
\eea
Introducing the bi-resolvent $W_c(z_1,z_2)= \langle \Tr(\frac{1}{z_1-\tilde{M}^{(c)}}) \Tr(\frac{1}{z_2-\tilde{M}^{(c)}})\rangle_{\text{cumulant}}$ and the resolvent $W_c(z) =  \frac{1}{N}\langle \Tr(\frac{1}{z-\tilde{M}^{(c)}}) \rangle$ 
and summing eq.\ref{eq:LeadSD} over $k$ weighted with a counting variable $z$ at the leading order in $1/N$:
\be
W_c(z)^2= (1-\alpha^2)zW_c(z)-(1-\alpha^2) .
\ee
\section{Conclusion}

In this paper we started the analysis of the $T^4$ melonic tensor model in any dimension $D>2$ using its matrix formulation and 
standard techniques used in the matrix models context. 
We have been able to compute the $2$-point function in the $N\rightarrow \infty$  and the result agrees with what have been known from earlier tensor models computation. Moreover we have been able to determine an Ansatz for the $1/N$ expansion of the eigenvalues leading to the calculation of their distribution at LO and NLO. As a result we discovered a collapsing of all eigenvalues at LO and the NLO is governed by a Wigner law's of width $\frac{1}{(1-\alpha^2)}$. This can be, in principle, continued further and should permit to obtain all the subsequent orders.
We conclude by a few questions. Can we determine more general relationships between observables of the tensor model and the one of its matrix model? Does the relation of the Proposition \ref{prop:TensMat} have a combinatorial interpretation? Can we relate the results analytically obtained here to combinatorics of the corresponding graphs of tensors? Can we use the matrix model written as in eq. \ref{eq:shiftmatrix} to investigate multiple scaling limits?
Is there any integrability property of this model (see \cite{TensGiv})? Does the Topological Recursion applies to this model?  

\section*{Acknowledgements}
Thanks are due to Vincent Rivasseau for proposing this work and following closely its progress. S. Dartois also acknowledges Valentin Bonzom for numerous discussions on matrix models and the matrix formulations of some tensor models. S. Dartois is partially supported by the ANR JCJC CombPhysMat2Tens grant. Bertrand Eynard thanks Centre de Recherches Math\'ematiques de Montr\'eal, the FQRNT grant from the Qu\'ebec government, Piotr Sulkowski and the ERC starting grant Fields-Knots. All authors are grateful to Laboratoire de Physique Th\'eorique d'Orsay for being an excellent working place.

 \noindent
{\small ${}^{a}${\it Laboratoire de Physique Th\'eorique,
Universit\'e Paris 11, 91405 Orsay Cedex, France, EU}} 
\\
{\small ${}^{b}${\it  LIPN, Institut Galil\'ee, CNRS UMR 7030, Universit\'e Paris 13, 
 F-93430, Villetaneuse, France, EU}}
\\
{\small ${}^{c}${\it IPhT, Institut de Physique Th\'eorique
Orme des Merisiers batiment 774, F-91191 Gif-sur-Yvette Cedex France, EU}} 
\\
{\small ${}^{d}${\it Centre de Recherches Math\'ematiques, Universit\'e de Montr\'eal, Montr\'eal, QC, Canada}
\\
{\small ${}^{e}$ {\it LAREMA, CNRS UMR 6093, Universit\'e d'Anger, D\'epartement de math\'ematiques, Facult\'e des Sciences, 2 Boulevard Lavoisier, 49045, Angers, France, EU}} 
\\
${}^{\ddagger}$ Author's e-mail: viet-anh.nguyen@polytechnique.edu, stephane.dartois@lipn.univ-paris13.fr, Bertrand.eynard@cea.fr

\end{document}